\documentclass[12pt,prd,amsmath,amssymb]{revtex4}% Physical Review 

 \usepackage{epsf}

\usepackage{amsmath,amsfonts}

\begin{document}

\title{Aspects of open strings in Rindler Space}

\author{David Berenstein$^\dagger$, Hee-Joong Chung $^{\dagger, \ddagger}$
}
\affiliation{$^\dagger$ Department of Physics, University of California at Santa Barbara, CA 93106\\
$^\ddagger$ Department of Physics, Yonsei University, Seoul, 120-749, Korea}

\begin{abstract}
We study open string configurations in Rindler space suspended from D-branes at finite distance from the Rindler horizon. This is a toy model for strings in the near horizon region of a black hole and has  applications for the study of strings 
configurations of heavy quarks in the AdS/CFT duals of hot field theories, as well as other applications to the study of open strings ending on giant gravitons. We find that this setup produces very similar results to those that have been found in the AdS black hole setup, but it is
much more tractable analytically. We also comment on some quantum applications of our studies to the understanding of the spectrum of strings ending on giant gravitons.

\end{abstract}

\maketitle

\section{Introduction}

An important part of the study of string background proceeds by suspending strings from brane configurations. These types of configurations have proved very useful in the study of Wilson loops in AdS/CFT \cite{Mwil,RY, RTY,BISY},  and have also been recently related to the problems of
energy loss of quarks in toy models of hot QCD \cite{HKKKY,G} (other calculations with different tools and results related to the energy loss of quarks were performed essentially simultaneously \cite{LRW1,CT}).

In the present paper we will study the problem of open strings in Rindler space that are suspended at finite distance from the Rindler horizon. Our motivation to study these configurations
arises from some observations made in \cite{BCV2}, where it was noted that in the study of strings suspended from a giant graviton that the ends of the string were accelerated and it was suggested that some of the peculiarities of the spectrum of these strings could be related to 
physics in a constant gravitational field. The idea that we want to explore was to set this toy model in a constantly accelerated frame in flat space. This is captured most easily by the study of the dynamics of open strings in Rindler space, where we provide the constant acceleration
at the ends of the string. This is, we suspend the strings from a brane at finite distance from the Rindler horizon. 

Since a Rindler wedge also describes the near horizon region of Scwarschild black holes, this
toy model we propose is also a limit of the studies of strings suspended at finite distance from an AdS black hole that have been used to study heavy quarks in a hot plasma, or finite temperature effects in AdS/CFT \cite{RTY,BISY,HKKKY,G}.
A general analysis for suspended strings has been performed in 
\cite{ASS}.
 What we find is a model that is analytically much more tractable than the study of strings in AdS, and where we can study in detail the stability of the string configurations.  After all, Rindler space is just a slice of flat space, and the closed string in flat space is solvable. With the open string, solvability depends on the boundary conditions. 

What we have found in our studies is that the open strings on Rindler space follow very closely 
the calculations made for strings suspended in an AdS black hole. The difference relies on the fact that our setup is much more tractable analytically and that the interesting string configurations can be described entirely in terms of hyperbolic functions. This also helps 
in the study of perturbations of the configurations.

This paper is organized as follows. In section \ref{sec:static} we study static Nambu-Goto strings suspended from branes at finite distance from the horizon and we solve the Nambu-Goto equations.  In section \ref{sec:energy} we calculate the energies of various static string configurations suspended from branes and we can argue in this way which configurations are globally stable. We can address in this way a phase diagram of configurations where we expect a first order transition. In section \ref{sec:stab} we study the problem of perturbative stability for configurations. This study gives a more detailed description of when the suspended string configurations disappear and get replaced by strings that get stretched to the horizon.
In section \ref{sec:vel} we consider string configurations moving at finite velocity. These string configurations are interesting as a limit of the drag force calculations in the AdS/CFT setup.
Later, in section \ref{sec:gg} we study the applications of our results to the problem of giant gravitons and we argue for various properties of the quantum spectrum of strings with accelerated endpoints. Finally, we close the paper with some concluding remarks.

\section{Static open string configurations in Rindler space}\label{sec:static}

Let us start by considering the following change of coordinates from flat space to Rindler coordinates.
We start with two coordinates $w,T$ with flat metric $ds^2=-dT^2+dw^2$. We now use the following parametrization of $T,w$:
\begin{eqnarray}
w&=& x \cosh (\kappa t)\\
T&=& x \sinh(\kappa t)
\end{eqnarray}
where we insist that $x>0$.
This coordinate map $t, x$ covers the region $w>|T|$ (the Rindler wedge). 
The parameter $\kappa$ is arbitrary. We want to keep it variable for our discussion. 
The metric in the new coordinate system is given by
\begin{equation}
ds^2 = - \kappa^2 x^2 dt^2 + dx^2
\end{equation}
Notice that the metric has a symmetry under translations of $t$. This corresponds to a boost symmetry in flat space.

In particular, one can consider the trajectory of a particle that stays at a fixed value of $x$.
This will be given by a hyperbola $w^2-T^2= x^2$. Near $T=0$, this gives us
\begin{equation}
w \sim x+\frac {T^2}{ 2x}
\end{equation}
so that the acceleration in natural units is $1/x$. Because of the boost symmetry, the intrinsic acceleration of the particle along this trajectory is constant and Rindler space serves as a toy model of a particle that is held in a constant gravitational field.

We should notice that there is a coordinate singularity where $x=0$. This is the Rindler horizon.
Since $g_{tt}$ depends on $x$, we can think of it as a non-zero gravitational potential that wants to attract  particles in the Rindler wedge towards the minimum of the potential, located at $x=0$. If a particle reaches $x=0$, it crosses the horizon and it's future evolution can not affect any particle whose trajectory lies completely within the Rindler wedge.

Rindler space is also a limit case of the geometry near a black hole horizon. 
Thus the physics of objects in the Rindler wedge is also used as a toy model for the behavior of these same objects near a black hole horizon. In particular, it is interesting to study 
how much we need to accelerate objects in order to keep them from falling. As explained above, as we make $x$ larger, we need less force to keep a particle at fixed distance from the horizon, since the intrinsic acceleration is smaller.

For example, if we have a very weak RR field, D-branes will be accelerated by the 
RR field, while strings will not feel the effects of the RR field. One can study the problem 
to first order in the RR field, before one has gravitational back-reaction to the RR field.
A D-brane can be put at a fixed distance from the Rindler horizon by this mechanism:
the RR field will sustain it in place. Strings, on the other hand, are not charged with respect to RR fields and they will fall.

Now, we want to consider a setup where a semi-classical open string is suspended from a pair of D-branes in Rindler space, where the D-branes are at fixed distance from the Rindler horizon. We have convinced ourselves that this is possible. There are various reasons to study this problem. We have explained these reasons in the introduction. Because our discussion will be (semi) classical, we are interested in studying just the Nambu-Goto string and we can ignore supersymmetry and RR backgrounds.

In order to fix the boundary conditions, we will consider that we have two D0-particles at fixed distance from the Rindler wedge. The particles will be separated along a third direction $y$, and they will be located at $x_1$ and $x_2$. If there is a static string configuration (with respect to $t$) with the given boundary conditions, symmetry suggests that the string will lie along the $x,y$ hyperplane. Thus, we can reduce the problem to the  study of a 
Nambu-Goto string in $2+1$ dimensions in the Rindler wedge geometry.

The full metric will be
\begin{eqnarray}
ds^2 = g_{\mu\nu}dX^{\mu}dX^{\nu} = -\kappa^{2} x^{2} dt^{2} + dx^{2} + dy^{2},\label{eq:Rindm}
\end{eqnarray}
where $\kappa$ is a surface gravity.

We will now use a worldsheet parametrization where($\tau, \sigma$) are parameters describing the worldsheet of  a string. We will choose the static gauge
\begin{eqnarray}
 t = \tau \\
 x(\tau,\sigma) = f(\sigma)\\
 y(\tau,\sigma) = \frac{d}{\pi} \sigma ~,
\end{eqnarray}
where $d$ is distance between two end points (this is the usual convention where $\sigma\in [0,\pi]$). Let the metric on the worldsheet be $\gamma_{\alpha\beta}$, this can be calculated by

\begin{eqnarray}
\gamma_{\alpha\beta} = g_{\mu\nu}\frac{\partial X^{\mu}}{\partial
\xi ^{\alpha}}\frac{\partial X^{\nu}}{\partial \xi^{\beta}} ~,
\end{eqnarray}
where $~\xi^{0}=\tau$, and $~\xi^{1}=\sigma$. \\

Then,
\begin{eqnarray}
\gamma_{\tau\tau}& =& g_{\mu\nu}\frac{\partial
X^{\mu}}{\partial\tau}\frac{\partial X^{\nu}}{\partial \tau} =
g_{00} = -\kappa^{2}x^{2} = -\kappa^{2}f^{2}\\
\gamma_{\tau\sigma} &=& 0\\
\gamma_{\sigma\sigma} &=& g_{\mu\nu}\frac{\partial X^{\mu}}{\partial
\sigma}\frac{\partial X^{\nu}}{\partial \sigma} = (\partial_{\sigma}f)^{ 2 } + \left(\frac{d}{\pi}\right)^2
\end{eqnarray}

The Nambu-Goto string action S is,
\begin{eqnarray}
S = - \frac{1}{2\pi\alpha'} \int d\tau d\sigma \sqrt{-\gamma} ~,
\end{eqnarray}
where $\alpha'$ is the slope parameter, and $\gamma$ is determinant of $\gamma_{\mu\nu}$. 

Then,
\begin{eqnarray}
S = - \frac{1}{2\pi\alpha'} \int d\tau d\sigma \sqrt{\kappa^{2}f^{2}[(\partial_{\sigma}f)^{2} + (d/\pi)^2]}\label{eq:action}
\end{eqnarray}

If we use instead $y=\frac{d}{\pi}\sigma +c$, where $c$ is a constant, and integrate over $t$,
we get that 
\begin{eqnarray}
S = -\frac{\kappa \Delta t }{2\pi\alpha'} \int dy f\sqrt{(\partial_{y}f)^{2}+1}
\end{eqnarray}

Since this is a static configuration, the total energy $E$ of the string is the quotient of the action by $\Delta t$ and we find
\begin{eqnarray}
E = \frac{\kappa}{2\pi\alpha'} \int dy f \sqrt{(\partial_{y}f)^{2} + 1}
\end{eqnarray}
notice that $\kappa$ factorizes. This is just the statement that $\kappa$ can be reset to any value we want by changing clocks (the units of time). The string configuration is independent of time, and therefore $\kappa$ is not relevant.

Now, we just want to minimize the energy with respect to $f$.

Because there is no explicit dependence on $y$ in the Energy density $f \sqrt{(\partial_{y}f)^{2} + 1}$, there is a conserved quantity for the variational problem we are considering (If one thinks of $y$ as a time variable for a dynamical system of a single variable, the associated conserved quantity would be the energy of the system.)
\begin{eqnarray}
\textit{constant} &=& \frac{- \partial (f \sqrt{(\partial_{y}f)^{2} + 1})}{\partial(\partial_{y}f)}\partial_{y}f + f \sqrt{(\partial_{y}f)^{2} + 1}\\
&=& - f \frac{(\partial_y f)^2}{\sqrt{(\partial f)^2 + 1}} + f \sqrt{(\partial_{y}f)^{2} + 1}\\
&=& \frac{f}{\sqrt{(\partial_{y}f)^2 + 1}}
\end{eqnarray} \\

Let us assume that there is a point of closest approach to the horizon, where $f=L$. We will label this point by $y=0$. This fixes the value of $c$ that we had introduced before.

Hence, we have reduced the second order differential equation problem to a first order differential equation
\begin{equation}
\frac {f}{\sqrt{(\partial_{y}f)^2 + 1}} = L \label{eq:Ldef}
\end{equation}
so that after some manipulations we get
\begin{equation}
dy = \frac{df}{\sqrt{(f/L)^2-1}}
\end{equation}
notice that in this problem the constant of integration and the height of closest approach agree.

From the last equation, we get the most general static solution to the variational problem.
For this we use the auxiliary variable $u = f/L$, so that $u=1$ at $y=0$.
\begin{eqnarray}
y = L \int_{1}^{ f/L} \frac{du}{\sqrt{u^{2} - 1}} \\
= L \ln( f/L + \sqrt{( f/L)^2 - 1}) 
\end{eqnarray}
and this can be inverted to find
\begin{equation}
 f = L \cosh(y/L) ,
\end{equation}

Notice that the second constant of integration was $c$. Here we have used a choice where 
$y=0$ is the point where $f$ is at a minimum. The most general solution of the second order differential equation would be
\begin{equation}
 f = L \cosh((y-c)/L) 
\end{equation}
In order to match the boundary conditions, we need to fit $f(y_1) = x_1$,
$f(y_2) = x_2$, with $y_2-y_1= d$. Since $f$ depends on two parameters, and we are matching $f$ at two values, having the three numbers $x_1, x_2, d$ is equivalent to
giving the three numbers $L, c, d$. We can dispense of $c$, by requiring that at $y=0$
$\partial_y f =0$, so that we can reparametrize our solutions in terms of fixed values 
of $L, y_1, y_2$.

Since $L$ is the place of closest approach to the horizon, $L$ measures the maximum acceleration that a piece of string is subjected to.

We should notice that in general one expects two different solutions for a static suspended string between two points $x_1, x_2$ at fixed distance $y$. If we choose the symmetric case where $x_1=x_2=x$ for simplicity, and we vary $L$, we get two catenary curves that intersect
at two points: when $L$ becomes smaller, the hyperbolic cosine grows faster with $y$, but it also gets closer to the horizon. This is why the two different curves intersect at two points. 
The two points of intersection vary as we vary $L$.

A plot of $L\cosh(y/L)$ for various values of $L$ is shown in the following figure (figure \ref{fig:curves}.) .

\begin{figure}[h]
\begin{center}
\epsfysize=5.8cm\epsffile{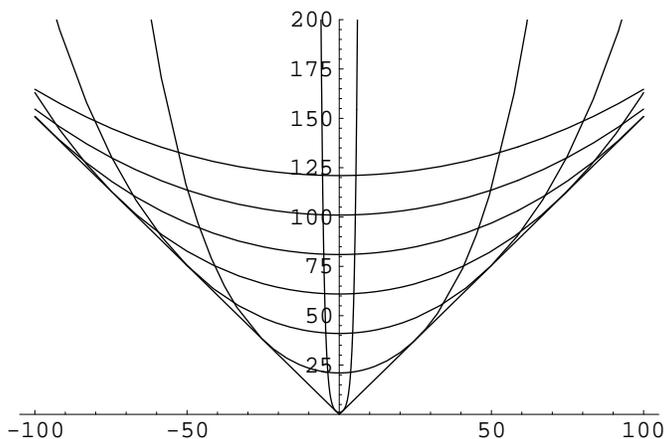}
{\caption{A plot of $L\cosh(y/L)$ for various values of $L$. We also show the tangent line to all the curves that passes through the origin.}\label{fig:curves}}
\end{center}
\end{figure}

The reader should notice that in the figure \ref{fig:curves} there is a region where 
$ y > s x $ that is not accessible to the solution found above. One can argue that this crossing point satisfies a  linear relation because if we scale $x,y$ by a factor of $r$, while leaving $t$ fixed,  
we get a dilatation of the metric. This can then be reabsorbed in the string tension, so that the classical problem one is trying to solve has scale invariance.

One can calculate $s^{-1}$ 
as the minimum value of $\cosh[y]/y$ (we have set $L=1$ for convenience). This is the slope of a line that crosses a  point on the curve and that also goes through the origin.We are looking for the tangent line to the curve. This requires us to solve for $\coth(y) = y$. This is easy to do numerically, finding a value 
of
\begin{equation}
y=1.19968 \label{eq:stabb}
\end{equation}
 and then for the region $1.50888 y> x$  there is no smooth static solution for a string suspended between two branes placed at the same distance $x$ from the
Rindler horizon. In particular, this value corresponds to 
$h/L= 1.81017$, where $h$ is the value of $x$ corresponding to this point on the curve.
The quantity $h/L$ can be used to distinguish two curves that reach $h$ at a given value of $y=d$. The one with the smallest value of $L$ is the closest one to the Rindler horizon and will have the largest value of $h/L$ (we could equally well have used $d/L$ to distinguish the two curves).

For us, it will be important to figure out for which values of $y, L$ the solutions are stable. In particular we have found generically two values of $L$, $L_s, L_l$ (small and large) that are compatible with the boundary conditions.
We will study this problem in detail in the symmetric case.

There is a third solution that we have missed so far, because of our choice of gauge. This is a collection of two strings that start from the two initial $x$ positions and stay at fixed value of $y$, but that go through the horizon in a straight line. These solutions correspond to a string suspended from the branes, part of which has crossed the horizon. The solution can be obtained as the limit curve when we take $L\to 0$, and when the tip of the suspended string touches the horizon, we cut the solution. The curve of closest approach in the figure suggests this interpretation.

In total we have three static solutions in some region, and only one static solution in the other region.
This suggests a simple effective potential representation of the three solutions in terms of a 
one parameter family of curves. The curve would have two minima and one maximum.

\begin{figure}[h]
\begin{center}
\epsfysize=5.8cm\epsffile{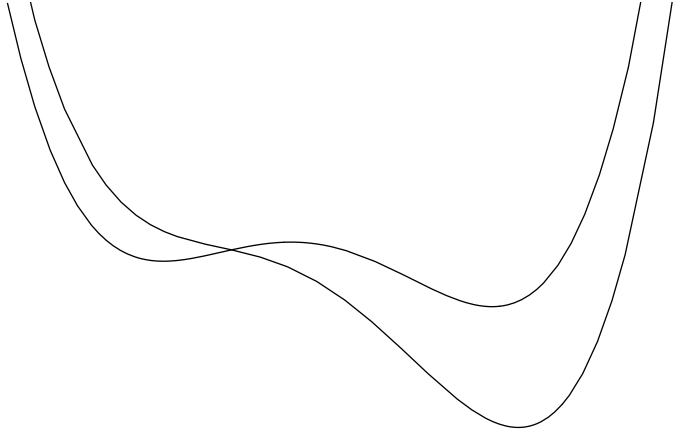}
{\caption{A pictorial representation in terms of an effective potential for a one parameter family of curves. Two values of $y$ are shown: one that leads to a curve with three extrema, and one that leads to a curve with a single extrema. \label{fig:pot}}}
\end{center}
\end{figure}

When we vary $y$, for $x$ fixed, we get different families of these curves, and eventually
the two minima and one maximum are replaced by a unique minimum. This is depicted in figure \ref{fig:pot}. We should find therefore that one of the smooth curves described above is classically stable under small pertrurbations, while the other curve is classically unstable. We would expect by continuity that the maximum is in between the two minima, so the curve that dips less towards the horizon should be stable, while the curve that dips closer to the horizon should be unstable. 
One also expects that when we change $y$, eventually the smooth curves disappear, 
and this should happen exactly at $1.50888 y=  x$ (this is where the two possible suspended string solutions become the same curve and is associated to having two extrema merge, a typical behavior near a phase transition).

Finally, the curve with the straight strings going to the horizon is classically stable (at least in the large y region). It becomes important to understand which of the two stable configurations has less energy. We expect that when we take $y\to 0$ at fixed $L$, the suspended string has smaller energy (it has smaller length), while close to the transition from three to one minimum, the curve with the straight strings has lower energy. 
If one can imagine thinking of $y$ and $x$ as thermodynamic variables, one would have a phase diagram where one would have a first order transition from the straight string phase to the suspended string at $y=b x$, where $b$ is calculable. 
This picture is very similar to the one discussed in \cite{FGMP}.

\section{Energy analysis} \label{sec:energy}

Suppose that there are two strings; one of them (string 1) is the string whose nearest point to horizon is L, and another configuration with two strings hanging straight and stretching towards the horizon (we will call this the string 2 configuration). Both strings share same end points, and the distance between two end points of strings is finite. \\

We calculate the energy difference $\Delta E$ between the two strings, and let's say that the end point is at $f = h$ (h is the maximal height that the string reaches). \\
For the first configuration (string 1),
\begin{eqnarray}
E_{1} &=& 2 \frac{\kappa}{2\pi\alpha'} \int dy f\sqrt{(\partial_{y}f)^2 + 1} \\
&=& \frac{4\kappa}{2\pi\alpha'} \int_{L}^{h} \frac{df}{\sqrt{(f/L)^2 - 1}} f \sqrt{(\partial_{y}f)^2 + 1} \\
&=& \frac{4\kappa}{2\pi\alpha'} \int_{L}^{h} \frac{df}{\sqrt{(f/L)^2 - 1}} \frac{f^2}{L} \\
&=& \frac{4\kappa L^2}{2\pi\alpha'} \int_{L}^{h} \frac{d(f/L)}{\sqrt{(f/L)^2 - 1}} \frac{f^2}{L^2} \\
&=& \frac{4\kappa L^2}{2\pi\alpha'} \int_{1}^{\bar u} \frac{u^2 du}{\sqrt{u^2 - 1}} ,\\
\end{eqnarray}
where $\bar u = h/L$.  \\

For string 2, the energy is $\frac{\kappa}{2\pi\alpha'}f^2$. Therefore total energy of the second configuration (string 2) is
\begin{eqnarray}
E_{2} = - 2 \int_{0}^{h} df \left(-\frac{\kappa}{2\pi\alpha'}2f\right)
=\frac{2\kappa L^{2}}{2\pi\alpha'} \int_{0}^{\bar u} 2u du
=\frac{2\kappa L^{2}}{2\pi\alpha'} \left[\int_{1}^{\bar u} 2u du + 1\right]
\end{eqnarray}
\\
Hence, energy difference $\Delta E = E_{1} - E_{2}$ is
\begin{eqnarray}
\Delta E = \frac{4\kappa L^2}{2\pi\alpha'} \left[\int_{1}^{\bar u} \left(\frac{u^2}{\sqrt{u^2 - 1}} - u\right) du - 1\right] \\
=\frac{\kappa L^2}{\pi\alpha'}\left[\bar u \sqrt{\bar u^2 - 1} + \ln(\bar u + \sqrt{\bar u^2 - 1}) - \bar u^2\right]\\
= \frac{\kappa L^2}{2\pi\alpha'}[(h/L) \sqrt{(h/L)^2 - 1} + \ln((h/L) + \sqrt{(h/L)^2 - 1}) - (h/L)^2]
\end{eqnarray}
\\
When $h \gg L$ we have,
\begin{eqnarray}
\Delta E \cong \frac{\kappa L^2}{2\pi\alpha'} [  \ln (2h/L)]
> 0
\end{eqnarray}
while when $h\to L$ the answer is negative.

This confirms the intuition we had expressed in the previous section. Obviously when $\Delta E<0$ a suspended string configuration has less energy than the pair of straight strings going to the bottom. The important point that we should notice is that for a fixed $L$, there is some value of $h$ where the second string configuration (two straight strings) have less energy than the suspended string.

In particular, for $h/L=1.81017$  we find a positive answer. This is the value where we calculated in the previous section for the tangent line to the curve that goes through the origin.
This means that where we expect the suspended string to stop existing as a static configuration, as argued in the previous section, we are in the regime where the straight string segments have lower energy. This is a consistency check of our intuition.

As a bonus to this analysis, we notice that when we have two different suspended strings from the same $h$, at least one of them has $h/L>1.81017$. Therefore it is always the case that one of the suspended strings has higher energy than the straight strings stretching to the horizon. This can be understood from the figure \ref{fig:curves}, where we notice that the intersection of two suspended strings lies in between the points where the curves touch the common tangent that crosses the origin.

A formal proof would have us take the function 
$$f(y)= \min( L_1 \cosh(y/L_1), L_2 \cosh[y/L_2]) - 1.5088 y$$
and notice that the function is continuous. Moreover, the function is positive and vanishes at two points (the two tangent points in the graph). Therefore it has a maximum in between these two points, and this maximum is the place where the two curves match (this is where one would get the discontinuity in the slope). This analysis confirms numerically the qualitative picture for the extrema that was described by figure \ref{fig:pot}. However, we still need to check that our conjectured unstable string is unstable under small perturbations exactly where we expect it to be.

\section{Stability analysis}\label{sec:stab}

In this section, we will look at linearized perturbations of the static solution to check for perturbative instabilities. This type of analysis has been done in AdS string in \cite{CG}, and 
for the string in an AdS black hole in \cite{FGMP}.

Let $\delta x(\tau,\sigma)$ be the perturbation, and $ f(\sigma)$ be the static solution to the equations of motion that we found, i.e. $f(y) = L \cosh (\frac{y}{L})$. We also have to impose that $\delta$ satisfy the appropriate boundary conditions so that the 
ends of the strings are fixed.

We now make the following ansatz for perturbations
\begin{eqnarray}
X^0 = t = \tau \\
X^1 = x(\tau, \sigma) =  f(\sigma) + \delta x(\tau,\sigma) \\
X^2 = y(\tau, \sigma) = \frac{d}{\pi} \sigma
\end{eqnarray}
with no perturbations in the other directions (it is easy to show that perturbations in directions transverse to the $x^0, x^1, x^2$ hyperplane are stable). We have used our gauge freedom to choose a convenient parametrization of the string. Thus the analysis reduces to a simple partial differential equation
for $\delta x(\tau, \sigma)$.

The Lagrangian of this perturbed string is $\mathcal {L} =  (-\gamma)^{\frac{1}{2}}$, and we expand the lagrangian to second order in $\delta$

\begin{eqnarray}
\gamma_{\tau\tau} &=& g_{\mu\nu} \frac{\partial X^{\mu}}{\partial \tau}\frac{\partial X^{\nu}}{\partial \tau} = g_{00} + g_{11}\delta \dot x \delta \dot x = - \kappa^2 ( f + \delta x)^2 + (\delta \dot x)^2 \\
\gamma_{\sigma\sigma} &=& g_{\mu\nu} \frac{\partial X^{\mu}}{\partial \sigma}\frac{\partial X^{\nu}}{\partial \sigma} = g_{11} \frac{\partial X^{1}}{\partial \sigma}\frac{\partial X^{1}}{\partial \sigma} + g_{22} \frac{\partial X^{2}}{\partial \sigma}\frac{\partial X^{2}}{\partial \sigma} = ( f' + \delta x')^2 + (d/\pi)^2 \\
\gamma_{\tau\sigma} &=& g_{\mu\nu}\frac{\partial X^{\mu}}{\partial \tau}\frac{\partial X^{\nu}}{\partial \sigma} = g_{11} \frac{\partial X^{1}}{\partial \tau}\frac{\partial X^{1}}{\partial \sigma} = \delta \dot x ( f' + \delta x')
\end{eqnarray}

Then,
\begin{eqnarray}
\mathcal {L} =[-(-\kappa^2(f + \delta x)^2+(\delta \dot x)^2)(( f' + \delta x')^2+(d/\pi)^2) + (\delta \dot x)^2 ( f' + \delta x')^2]^\frac{1}{2} \\
= [(\kappa^2( f + \delta x)^2 - (\delta \dot x)^2)(d/\pi)^2 + \kappa^2( f + \delta x)^2(\bar f' + \delta x')^2]^\frac{1}{2}
\end{eqnarray}
and this can be readily expanded. After some algebra, we find that 
the equations of motion for the perturbation can be written as
\begin{eqnarray}
\frac {L^2}{\kappa ^2} \sinh^2(y/L)\delta \ddot x = L^4 \cosh^2(y/L)\delta x'' - 2L^3\cosh(y/L)\sinh(y/L)\delta x' + L^2\cosh^2(y/L)\delta x
\end{eqnarray}

Let us use the variable $u = y/L$, differentiation is by $u$, which is $\frac{\partial}{\partial \sigma} = L\frac{\partial}{\partial y}$. 

Thus the  equation that needs to be analyzed  becomes
\begin{eqnarray}
\frac {1}{\kappa ^2} \tanh^2 u ~ \delta \ddot x = \delta x'' - 2\tanh u ~ \delta x' + \delta x
\end{eqnarray}

We can separate variables in the time direction trivially, by setting 
\begin{equation}
\delta x (\tau, u) = A(u)(e^{i\omega \tau}) +\bar A(u) e^{-i\omega \tau}
\end{equation} 
Since
$\delta x$ has to be real, we find that $\omega$ is either real or purely imaginary.
We thus find that $A(u)$ satisfies

\begin{eqnarray}
A''(u) - 2 \tanh u A'(u) + \left(1 + \frac{\omega^2}{\kappa^2} \tanh^2 u \right) A(u) = 0
\end{eqnarray}
Then general solution of this differential equation is,
\begin{eqnarray}
A(u) = c_{1}\cosh u ~P\left[\frac{1}{2}(-1+\sqrt{9-4(\omega/\kappa)^2}),i\omega/\kappa,\tanh u \right] \\ +c_{2}\cosh u ~ Q\left[\frac{1}{2}(-1+\sqrt{9-4(\omega/\kappa)^2}),i\omega/\kappa,\tanh u \right],
\end{eqnarray}
where $P[\nu,\mu,x] = P^{\mu}_{\nu}(x)$, which is Legendre function of the first kind, and $Q[\nu,\mu,x] = Q^{\mu}_{\nu}(x)$, which is Legendre function of the second kind. 
Written in this way,  this is not a particularly illuminating way to analyze the stability of the configurations. However, we should point out that the problem is in principle solvable in terms of 
known special functions. This is an improvement over the analysis in \cite{FGMP} that required
solving the problem numerically.

For the analysis we find it convenient to set $A(u) = \psi (u) \cosh u $, and rewrite the differential equation for $\psi(u)$,
\begin{eqnarray}
\psi''(u) + (2 + ((\omega/\kappa)^2-2)\tanh^2 u ) \psi (u) = 0
\end{eqnarray}

In a more usual Schr\"{o}dinger equation form, it is written as
\begin{eqnarray}
-\psi''(u) + [(2-(\omega/\kappa)^2)\tanh^2u] \psi (u) = 2 \psi (u)
\end{eqnarray}
In this equation, we can interpret $(2-(\omega/\kappa)^2)\tanh^2u$ as the potential for a one dimensional quantum mechanical problem, and the value of $2$ in the right hand side as the energy of the corresponding state.

Since the maximum value of $\tanh^2 u$ is less than 1 (and it attains this value at infinity), we find that for real $\omega>0$ the total associated energy is always larger than the asymptotic value of the potential (we are ignoring the boundary conditions for the moment). Under these circumstances, the wave function $\psi$ always has nodes at some finite value of $u$ and it asymptotes to a plane wave at $+\infty$ and $-\infty$, because we are in the continuum part of the spectrum.
Indeed, even for $\omega=0$ we find two zero's, although for $\omega=0$ the wave function is not normalizable in the $(-\infty, \infty)$ region (not even in the $\delta$ function 
normalization).
 
For $\omega=0$, in the one-dimensional Schrodinger problem we are exactly at energy equal to two, which is also the asymptotic value of the potential. 
One can then solve the differential equation, and for simplicity we use an 
even function. 
For $\omega=0$ the function $\psi$ is proportional to
\begin{equation}
\psi(u) \sim \frac{1}{2}\log\left(\frac{1+\tanh(u)}{1-\tanh(u)}\right)\tanh(u)-1
\end{equation}
and this equation has a zero close to $u=1.19968$.
Notice that this number is the same at which we argued in equation \ref{eq:stabb} that the instability appears. This can be thought of as a consistency check of our calculations.
This is also consistent with the observation of \cite{AAS} that instabilities appear always at these degenerate loci.

What we find is that for separations of $d/(2L)<1.19968$ the function above has no nodes in the appropriate interval, and it does not satisfy the boundary conditions at $y=d/2$. One can then increase $\omega$ until such an even solution appears. This value $\omega_0$ is the lowest frequency of oscillations. Thus we have that for $d/(2L)<1.19968$ the string is perturbatively stable.

Now we can consider the case that $\omega= i\Omega$ is imaginary. We find that 
the effective potential is now given by $(2+\Omega^2/\kappa^2) \tanh^2(u)$, and that the effective energy of the particle in the Schrodinger problem is $2$ which is less than the asymptotic value of the potential. Thus it resembles a bound state in the infinite line problem. Since the potential is attractive, there is always at least one normalizable bound state in the $(-\infty, \infty)$ region. Moreover, since at $\Omega=0$ we have nodes inside the region where we are trying to solve the problem, we find that there is at least one state that will solve the boundary conditions that we need for $\Omega>0$. This is, the string is necessarily unstable. 

The possible values of $\Omega$ we get are discrete because we also need to satisfy the boundary conditions at $y=\pm d$.
Also, if $\Omega$ is made very large, we can expand the potential around $u=0$ (the minimum), finding an approximate harmonic oscillator schrodinger problem, of frequency proportional to $\Omega$ in the large $\Omega$ limit.
If the energy $E=2$ is below the lowest energy of the associated  harmonic oscillator, then there is no solution that satisfies the differential equation. This means that the unstable modes have bounded exponential growth and there is only finitely many of them. 

One can argue that this bound is related to the fact that the instability is due to a gravitational
pull of the string towards the Rindler horizon, and that a string that starts as a small perturbation of a static configuration will not accelerate faster than what the acceleration of free fall can provide.

To summarize, we have found that for fixed $L$, if the separation of the ends of the suspended string are too large, then the string is unstable under small perturbations, whereas if the separation is small enough, it is stable under small perturbations. We also showed that 
the string is marginally unstable at a particular separation of the ends. This matched exactly the intuition and numerical value that we had obtained from studying the static solutions in section \ref{sec:static}.

\section{Adding velocity}\label{sec:vel}

Now we will consider the same problem as in the beginning, a string suspended from D-branes in Rindler space, but we will have the D-branes stay at fixed distance from the horizon and we will ask that the ends of the string move at constant speed $v$ in some direction $z$, which will also be perpendicular to the separation $y$ between the two ends of the string. This is the near horizon limit of the problem studied in \cite{CGG,AEV}. The case of rotating strings was studied in \cite{PSZ}.

Indeed, the motion at constant velocity is natural for a point particle restricted to the wolrdvolume of the D-branes that are at fixed distance from the Rindler horizon. This is because on the D-brane worldvolume, the induced metric 
from equation (\ref{eq:Rindm}) is flat in a natural flat coordinate basis. Free particles in such a metric move at constant speed. Requiring the strings to move at constant speed seems like a natural generalization of this observation.

We will first consider static constant velocity strings with a  profile given by
\begin{eqnarray}
t &= &\kappa^{-1}\tau\\
x&=& f(\sigma)\\
y &=& \frac d\pi\sigma\\
z &=& v \tau +g(\sigma)
\end{eqnarray}
where we have made use of reparametrization invariance to write the coordinates in a gauge similar to static gauge.
After writing the Nambu-Goto action and looking at the equations of motion we find that the equations of motion for $g$ are solved for $g$ constant. This is the same result found in \cite{CGG}, but in a specialized case.

 This simplifies the problem substantially. Using this information we find that the problem we have to solve is a minor modification of the one we encountered in equation (\ref{eq:action})
\begin{equation}
S = \frac {-1}{2\pi \alpha'}\int d^2\sigma \sqrt { (f^2-v^2)[(\partial_\sigma f)^2+(d/\pi)^2]}
\end{equation}
A straightforward analysis shows that there is a first integral of motion of the form
\begin{equation}
\frac{\sqrt{f^2-v^2}}{\sqrt{(\partial_y f)^2+1}} = L
\end{equation}
a minor modification of equation (\ref{eq:Ldef}). At the minimum of the string we have $\partial_y f|_{y_0}=0$, and this lest us solve  $f(y_0) =\sqrt{L^2+v^2}$. One can argue in general that $f(y_0)\geq v$ from having the correct signature of the induced metric on the string worldsheet.

 Again,  we can integrate the equation to find the solution
\begin{equation}
f_{L,v}(y) = (L^2+v^2)^{1/2} \cosh(y/L)
\end{equation}
In conventions where the point of closest approach is at $f(y_0)= H$, we have
\begin{equation}
f_{H,v}(y) = H \cosh(y/\sqrt{H^2-v^2})
\end{equation}

This solution has the same qualitative features than the solutions found before. Again, one can argue from continuity of what we have discussed so far that if two strings have the same endpoints, then the closest string to the horizon is unstable. However, the normal mode analysis will now mix $z$ and $x$ perturbations. We will not perform this analysis for this case.

What is more interesting to study is the solution that corresponds to the two straight strings hanging towards the horizon, but now with velocity $v$. 

The corresponding ansatz for the Nambu-Goto string is that the coordinate $y$ is constant.
As such, the naive ansatz for static gauge that we have been using is not valid. Instead, we need to consider a gauge of the form
\begin{eqnarray}
t&=&\kappa^{-1} \tau\\
y&=&0\\
x&=& \sigma\\
z&=& v\tau +g(\sigma) \label{eq:ansstr}
\end{eqnarray}
The induced metric is 
\begin{equation}
g_{ind} \sim \begin{pmatrix} -\sigma^2+v^2 & v \partial_\sigma g\\
v\partial_\sigma g& 1+ (\partial_\sigma g)^2
\end{pmatrix}
\end{equation}

Again, as above, there is a solution to these equations for $g=0$. However, when such a string hits the surface $x=v$, the induced metric changes signature, and the full string solution is argued to be unphysical. 

If we write the parametrization of this string in regular flat space coordinates as opposed to the Rindler coordinates, we find that
\begin{eqnarray}
T&=& \sigma\sinh(\tau)\\
w&=& \sigma \cosh(\tau)\\
Z&=& v\tau
\end{eqnarray}
in the same $T,w$ coordinates of section \ref{sec:static}.
Since $\sigma>v$, we can do a change of variables and replace $\sigma = v \cosh \tilde \sigma$. 
We find this way that the straight string in Rindler coordinates is parametrized as 
\begin{eqnarray}
T&=& v \cosh(\tilde \sigma)\sinh(\tau)\\
w&=& v \cosh(\tilde \sigma) \cosh(\tau)\\
Z&=& v\tau
\end{eqnarray}
and this is a double analytic continuation of the standard rotating folded string solution that gives rise to the leading classical Regge trajectory.
We do the double analytic continuation of  the solution by taking
$\tau \to i \tau$ and $\tilde \sigma\to i \tilde \sigma$,  so that we find
\begin{eqnarray}
\tilde T&=& i v \tau\\
\tilde w &=& v \cos(\sigma) \cos(\tau)\\
\tilde Z&=&  iv \cos(\sigma)\sin(\tau)
\end{eqnarray}
We can understand this process as changing the sign of the metric on the worldsheet (exchanging space and time). We notice that $\tilde T$ and $\tilde Z$ are imaginary, so we can 
also change space and time in the target space (thereby making $i\tilde T=Z'$ and
$i \tilde Z = T'$), and we can compare with the folded string solution
\begin{eqnarray}
T'&=& \tau\\
w'&=& R \cos(\sigma/R) \cos(\tau/R)\\
Z'&=&  R \cos(\sigma/R)\sin(\tau/R)
\end{eqnarray}
finding the same functional form.
This parametrization of the folded string is the one for static gauge and should not be confused with the conformal gauge parametrization.

Here, it becomes obvious that the point on the worldsheet that corresponds to $\sigma=0$ is traveling at the speed of light and is to be interpreted as a cusp, just like the folding point of the folded string solution. 

A second solution can be constructed  where we can choose $g$ to be smooth suggests the following gauge
\begin{eqnarray}
t&=& \kappa^{-1} \tau\\
y&=&0\\
x&=& f(\sigma)\\
z&=& v\tau +  \sigma
\end{eqnarray}
This second ansatz gives us the following action
\begin{equation}
S \sim - \int d^2\sigma \sqrt{g_{ind}} = \int \sqrt{f^2+ f^2 \partial_\sigma f^2 - v^2 \partial_\sigma f^2}
\end{equation}
It is easy to find the first integral of motion,
\begin{equation}
\frac{f^4}{f^2+(f^2-v^2)( \partial_\sigma f)^2} = G
\end{equation}
The equation for motion of $f$ can then be integrated in terms of an Appell hypergeometric function, but it is hard to invert these formulas to find $f$ itself. This solution is similar
to the one found in \cite{HKKKY,G}. This solution requires a force dragging the string, and the string has an infinite energy.

We now want to argue that the string should have the property that 
$\partial_\sigma f $ is finite at $x=v$. 

Now, assuming $\partial_\sigma f$ is finite at $f=v$, gives us $G= v^2$. Notice that for general solutions $\partial_\sigma f$ can not vanish, so we find that for other values of $G$ it must be the case that $\partial_\sigma f|_{f=v}=\infty$ if the string reaches the $x=v$ region at all.

From $G=v^2$ we get that the problem simplifies and we have to solve
\begin{equation}
f^2  = (\partial_\sigma f) ^2 v^2
\end{equation}
This can be readily integrated to find that 
\begin{equation}
f =   v \exp (\pm \sigma /v)
\end{equation}
where we have chosen $\sigma=0$ to represent the point $x= v$. In this solution we need to choose the sign. However, only one solution is physical 
in that the suspended string is trailing behind the motion of the end of the string on the D-brane.
Notice also that in the limit $v\to 0$ the coefficient of the exponential becomes infinite and we recover the vertical string configuration.

What we have found mimics exactly, in a simpler setting, the solutions found in \cite{HKKKY,G}. 
 Also, as is clear from our analysis of the stability of the static string in Rindler space, the analysis of the stability of these types of solutions should be a lot simpler than in the AdS black hole geometries.

\section{Applications to giant gravitons}\label{sec:gg}

Part of the initial motivation to do this study, was to get a simplified model 
of open strings ending on giant gravitons, and in particular in the case of open strings attached to non-maximal giant gravitons in $AdS_5\times S^5$. In the study of such strings, via the AdS/CFT correspondence, one can consider the problem of calculating the energies
of open strings via a dual perturbative gauge theory computation. The gauge theory dual to this proposal for general giant gravitons was studied first in \cite{BCV}. This gave rise to a spin chain model with variable numbers of sites, and after a bosonization transformation, it could be understood as a chain of Q-bosons with non-diagonal boundary conditions.

The one boson site model was solved in 
\cite{BCV2}, where it was found that the spectrum consisted of a single bound state and a gap to a continuum band of states. Moreover, it was found that the spectrum of the finite size 
spin chain was continuous, and this indicated that the classical string solution should experience some runaway behavior where the string grows until non-planar effects become important.
One of the important questions is whether the problem is integrable or not.
This was further studied  in \cite{AO} for the case of maximal giants. The main difficulty in finding integrability is that the continuous spectrum prevents a solution in terms of a Bethe Ansatz. Recently, remarkable progress has been made in this direction in the case of 
the closed string on $AdS_5\times S^5$ in the work \cite{BES}.

It was observed that for a general open string attached to these giant gravitons that since the giant gravitons are affected by RR fields, while the strings are not, that the motion of the giant gravitons is not geodesic and that the ends of the string experience a similar effect as that of being in an accelerated frame. The simplest such model would happen in flat space, with D-branes at constant acceleration relative to an inertial frame. If the reason for the open strings in AdS to show a continuum is 
due to the acceleration of the ends of the string, we should be able to reproduce the same effects with the open strings in Rindler space.

What we want to do now is to qualitatively explain some of the features that were found in
those papers within the context of the toy model we are analyzing.

The first surprise, that was explicitly computed in \cite{BCV}, was that the spectrum of states
of a "single site" spin chain could have one normalizable ground state and all the other states were in the continuum.

As we saw in the description of the strings suspended in Rindler space, we had some configurations that were stable strings, some were unstable strings, and some have strings that stretched all the way to the horizon. 

Within this model, one has a discrete spectrum of states near the stable strings (these can be metastable if the energy is not sufficiently low to be below the stright string configuration) and for the strings stretching to the horizon we have a continuum spectrum of excitations. This is the familiar statement that we should use 
boundary conditions so that all excitations fall trough the horizon.
Since these give rise to quasinormal modes with complex frequency, one naturally associates this sector with a continuum spectrum. 

\begin{figure}[h]
\begin{center}
\epsfysize=5.8cm\epsffile{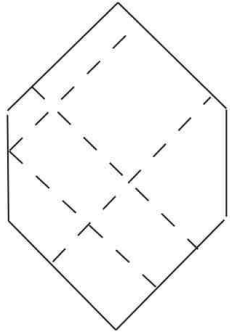}
{\caption{A possible causal structure for an open string worldsheet. Various null lines are shown as dashed lines. Notice that a line can bounce from the boundary at most once in this particular case. \label{fig:causal}}}
\end{center}
\end{figure}

Indeed, another explanation for the absence of a discrete spectrum can be obtained by understanding that in $1+1$ lorentzian geometries, the causal structure of an open string configuration is not always that of an infinite strip, and instead it can have regions  associated to asymptotic boundaries. This is illustrated in the figure \ref{fig:causal}.

 Indeed, in a semiclassical treatment of the study of fluctuations for a strip-shaped geometry one can argue that left and right-moving excitations reflect from the walls
of the open  string infinitely many times and one can apply a semiclassical quantization condition that produces a discrete spectrum around a static configuration.

Consider now the open string worldsheet associated to having boundaries at $\tau^2-(\sigma+L)^2= - a^2$, where we cut the region to the left of $(\tau,\sigma)=(0,-L-a)$, and to $\tau^2-(\sigma-L)^2=- a^2$,
where we cut the region to the right of $\sigma=L+a$. As we can see, in this configuration a left or right moving excitation can bounce at most once from the boundary. This is similar to the causal structure depicted in figure \ref{fig:causal}. Thus, there is no semiclassical action angle
variables that are periodic (bounce infinitely many times) and this prevents the system from being able to be described by 
a discrete set of frequencies. This is the origin of the continuous spectrum, even for open strings that are attached to a D-brane. It is easy to notice that this causal structure is not conformally equivalent to a strip. If we choose an isometric embedding of this structure in a spacetime, the ends of the string
must be accelerated with respect to each other in order for them not to be in causal contact. 
This requires us to look at systems that can accelerate the ends of the strings and make the string arbitrarily long at late times. This can not happen with static brane configurations in flat space, as one would have a conserved energy on the string that prevents it from getting arbitrarily long.

Now let us consider the configuration of strings that was studied in giant gravitons in \cite{BCV}. 
The limit of one site in the spin chain model corresponds to a short string. The ground state is a massless excitation of the D-brane, so the string is traveling at the speed of light. However, when we consider excitations of this string, we will be traveling close to the speed of light.

We should translate this into a short string close to the horizon, thus $y<x$, and $|x|\sim \ell_s$. This is, the string should be short and at distance of order the string scale to the horizon. If we add the velocity $v$ like we did in the previous section, the string should be short and very close to the "velocity of light" surface. 

Now we can ask if the spectrum will exhibit the features of the spin chain that were found in \cite{BCV}:
a ground state and a gap to a continuum. We find that for a short string, the frequencies of oscillations of the 
quadratic perturbations do not depend on the string tension. This is true in general for extended brane configurations, and the string is such an example. The frequencies only depend on the geometric profile of the string.  This is the statement that classically the string tension only serves to normalize the action and the energy, but without $\hbar$ there are no natural units for the frequencies.

However, in the quantum case, when we have $\hbar\neq 0$, the total energy of a single harmonic excitation of frequency $\omega$ is $\hbar \omega$, and this can be much higher than
the difference in energy between the stable and the unstable string, that does depend on the string scale $\ell_s$. We thus find that it is possible to have a stable ground state and to reach the continuum on the first excited state.

If we make the string longer (or if we increase the value of $\alpha'$) we can get a situation where the string has more bound states below the continuum line. In the spin chain model, this is adding more than one spin site to the chain.

Another question that has attracted attention is whether in general the smaller giant gravitons correspond to integrable boundary conditions on the string or not. Here, we do not have very much to say explicitly. We notice that since the problem of perturbations around a 
simple configuration was solvable in terms of known special functions, it is a hint that the boundary conditions we have studied in this paper might be integrable. We can also point out that the boundary condition that we have studied can be written as a hyperbole
\begin{equation}
(x^1)^2-(x^0)^2 = a^2
\end{equation}
This condition is very similar to the one studied for the circular brane in \cite{LZ}, that is closely related to the paperclip model \cite{LVZ}, where they gave strong evidence for integrability. The situation here is a natural analytic continuation of the circular brane. Moreover, one can argue that there are no winding modes on the hyperbole, so the theta angle of the analytically continued model should be zero. This deserves to be investigated further, but it is beyond the scope of the present paper.

\section{Conclusion}

In this paper we have shown that it is possible to analyze the problem of strings suspended from a D-brane at finite distance from the Rindler horizon in a lot of detail for various setups. We have made an analysis of the basic stability conditions for static suspended strings at zero velocity. We were able to show that the exact analytical calculations matches the geometric intuition for when the instabilities set in. Overall, this is very similar to the calculations done in
\cite{FGMP}.

 The problem we studied is a universal near horizon limit for strings suspended from branes in various black hole geometries.  The advantage of working in Rindler space is that the string is essentially in flat space and therefore is solvable, except perhaps for the boundary conditions. We have argued also that the boundary condition of being at finite distance from the Rindler horizon seems to be integrable. 
  
 One should also be able to consider general mesons that correspond to rotating strings (like the leading Regge trajectories), a problem that has been also analyzed in \cite{PSZ}. We think it would also be instructive to analyze configurations like the ones studied in \cite{LRW}. 
 
In general, for some of these problems of strings in AdS black holes it has been necessary to solve the Nambu-Goto equations numerically. If not at the level of the static suspended string, then at the level of the perturbations of a given solution.

In the simplified model we have studied the string seems to be completely solvable and this raises the possibility that one can find an exact analytical treatment for all of the interesting configurations that have been explored in the literature. It is also likely that one can find exact solutions of the strings being slowed down as their tails fall into the horizon. This would mean that in principle one can find a complete analytical solution for how unstable strings evolve, especially their long time evolution where the string perturbations become irrelevant.

Although this study is beyond the scope of the present paper, it would be very interesting to have these solutions. In particular this might be useful to understand analytically how quark bound states dissociate in a quark-gluon plasma.

\section*{Acknowledgements}

D.B. would like to thank C. Herzog, G. Horowitz, P. Kovtun and J. Polchinski for various discussions and correspondence related to this project. Work supported in part by  DOE, under grant DE-FG02-91ER40618. H.C. would like to thank UCSB for hospitality during his exchange student program, and the Division of International Education and Exchange of Yonsei University for providing him an opportunity to be an exchange student at UCSB.

\end{document}